\documentclass[12pt]{article}
\usepackage{amsmath}
\usepackage{amscd}
\begin{document}

\newcommand{\R}{{\bf R}}
\newcommand{\C}{{\bf C}}
\newcommand{\Z}{{\bf Z}}

\rightline{CERN-TH/2000-048}
\rightline{February 2000}
\vskip 2.5cm

  \centerline{\Large \bf Some Aspects of Deformations of}
\bigskip
\centerline {\Large\bf  Supersymmetric Field Theories}

\vskip 2.5cm

\centerline{S. Ferrara$^*$ and M. A. Lled\'o$^\dagger$}
\bigskip
\bigskip

\centerline{\it $^*$ CERN, Theory Division, CH 1211 Geneva 23,
Switzerland.} 
\bigskip
\centerline{\it $^\dagger$ Dipartimento di Fisica, Politecnico di Torino,}
\centerline{\it Corso Duca degli Abruzzi 24, I-10129 Torino, Italy, and}
\centerline{\it INFN, Sezione di Torino. Italy.}
\vskip 2cm
\begin{abstract}
  We investigate some aspects of Moyal-Weyl deformations of superspace and
their compatibility
 with supersymmetry. For the simplest case, when only bosonic coordinates
are deformed,
we consider a four dimensional supersymmetric field theory which is the
deformation of the 
Wess-Zumino renormalizable theory of a chiral superfield. We then consider
the deformation of 
a free theory of  an abelian vector multiplet, which is a non commutative
version of the rank 1 
Yang-Mills theory.  We finally  give the supersymmetric version of the 
$\alpha' \mapsto 0$ limit of the Born-Infeld action with a $B$-field
turned on,
 which is believed to be related to the  non commutative U(1) gauge
theory.

\end{abstract}

\vfill\eject
\section{Introduction.}

Deformations of mathematical structures have been used at different
moments in physics. 
When Galilean transformations between inertial systems were seen not to 
describe adequately the physical world, a deformation of the group law
 arose as the solution to this paradox. The Lorenz group
 is a deformation of the Galilei group in terms of the parameter
$\frac{1}{c}$. From the
mathematical point of view it is not difficult to imagine the deformation
of a group
inside the category of groups, but form the physical point of view it has
enormous
consequences. In this deformation scheme, the old structure is seen as a
limit or 
contraction when the parameter takes a preferred value.

The mathematical structure of quantum mechanics has also an ingredient of 
deformation with respect to classical mechanics. The first star product
or formal deformation of the commutative algebra of classical observables
was written in 
Ref.\cite{m}. The star product is a product in the space of formal series
in $\hbar$         
whose coefficients are   functions on the  phase space. It  is homomorphic
to the product of 
operators in quantum mechanics. In this case the deformation occurs inside 
 the category of algebras, although giving up the commutativity. 
 More complicated features arise in the mathematical framework of quantum
mechanics, and 
the first thing one can realize is that this deformation in the parameter
$\hbar$ is a formal
 deformation (that is, the series in $\hbar$ of a product of two functions
is not convergent),
unless strong restrictions are made on the  functions. Nevertheless, 
Bayen, Flato, Fronsdal, Lichnerowicz and Sternheimer realized in their
seminal papers \cite{bffls}
that a first approach to a quantum system could be studying  the formal
deformations of 
the classical one, leaving aside problems of convergence and of
construction of the Hilbert space.
The existence and  uniqueness of this deformation (up to gauge
transformations), finally showed in
\cite{ko} for any Poisson manifold,  supports the belief that the formal
deformation encloses 
the essential information of the quantum system. 
 
Much more recently, non commutative geometry has entered in physics in
different 
contexts. One context is string theory and M-theory. In their pioneering
paper, Connes, Douglas 
and Schwarz \cite{cds} introduced non commutative spaces  (tori) as
possible 
compactification manifolds of space-time. Non commutative geometry  arises
as a possible
scenario for short-distance behaviour of physical theories. Since non
commutative 
geometry generalises standard geometry in using a non commutative algebra
of 
``functions'', it is naturally related to the simpler context of
deformation theory.

Matrix theory \cite{bfss, ikkt}
 on non commutative tori is 
  related 
to 11 dimensional supergravity toroidal compactifications with 3-form
backgrounds.
In this framework, T-duality arises as Morita equivalence in non
commutative geometry
\cite{rs}. This lead to subsequent developments of Yang-Mills theories on
non commutative 
tori \cite{mz}, as the  study of their BPS states \cite{bm} and a
reformulation of T and U dualities
of Born-Infeld actions on non commutative tori \cite{hv}. Non commutative
geometry also appeared
in the framework of open string theory \cite{dh}.

 More recently, Seiberg and Witten \cite{sw}
identified limits in which the entire string dynamics, in
presence of a $B$-field, is described by a 
 deformed  gauge theory in terms of a Weyl-Moyal star product on
space-time.
In particular, they showed  that the pure quadratic gauge theory with
deformed abelian 
gauge symmetry is related through a change of variables to a non
polynomial gauge theory 
with undeformed gauge group. This brings a connection between the
Born-Infeld action and
gauge theories in non commutative spaces. In view of the fact that the
supersymmetric 
Born-Infeld \cite{cf} action naturally arises as the Goldstone action of
$N=2$ supersymmetry,
partially broken to $N=1$ \cite{bg}, it must be the case that this
interpretation should have its 
counterpart in the framework of deformed gauge theory. This connection
will be clarified in this paper. 
 Subsequently,
this  deformation of space-time was used  for  ordinary four dimensional
field theories with a 
$B$-field of maximal rank in $\R^4$ space-time.
It was shown that the deformed theories enjoy unsuspected renormalization
properties as well
as UV/IR connection reminiscent of string theory \cite{mrs}. 

Other approaches connecting deformation theory to theories of gravity have
also  
appeared in the literature.  Among others, we can mention the deformation quantization
of M-theory
\cite{mn}, quantum anti de Sitter space-time \cite{hs},  q-gravity
\cite{finki} and
gauge theories of quantum groups \cite{c}.

In this paper some issues related to  theories formulated in 
deformations of superspace are investigated. Aspects of non commutative
supergeometry \cite{ cz,s}
 and noncommutative supersymmetric field theories \cite{cr, gp, ss}
were recently considered in the literature. Section 2 is  a
self explanatory
 account of the Moyal-Weyl deformation generalised to superspace. In
particular we
 show that the 
deformation of a Grassmann algebra obtained with a Weyl ordering rule is a
Clifford algebra once we specify a value for the parameter (non formal
deformation). 
In section 3 we consider the compatibility of this deformation with the
supertranslation
group. In section 4 we present the simplest example of a deformed
supersymmetric field
theory, the Wess-Zumino model and we give its explicit expression in terms
of field
components. In section 5 we consider deformed gauge groups in superspace
and derive a non 
commutative version of rank 1 Yang-Mills theory which is then coupled to
chiral superfields. 
We show in particular, to first  order in the deformation parameter,
 that the change of
variables of Seiberg and Witten
 \cite{sw} to convert the rank 1 non commutative quadratic gauge theory to
a commutative 
higher derivative theory is consistent with supertranslations. 
Rank 1 $N$-extended 
super Yang-Mills theories are invariant under shifts of the $N$  gauginos
by $N$ constant,
anticommuting parameters, so they  can be regarded as Goldstone actions of
$2N$
supersymmetries spontaneously broken to $N$ supersymmetries.
 Finally in section 6 we derive the $\alpha'\mapsto 0$ limit of
supersymmetric
Born-Infeld actions, which are the starting point for comparisons with 
super Yang-Mills theories in non commutative superspaces.

\section{Star product in superspace.}

\subsection{Super Poisson bracket.}

A super vector space over $\R$ or  $\C$ is a $\Z_2$-graded vector space 
$V=V_0\oplus V_1$ with 
grading $p=0,1$. On $V$ we can give a associative operation $\cdot
:V\otimes 
V\mapsto V$  respecting the grading, that is
$$
p(a\cdot b)=p(a)+p(b).
$$
Then $V$ is a super algebra. A super Lie algebra is a super vector space
$V$
 with a bracket $[\;,\;]:V\otimes V \mapsto V$ satisfying
\begin{equation}
[X,Y]=-(-1)^{p_Xp_Y}[Y,X],\label{sp}
\end{equation}
\begin{equation}
[X,[Y,Z]]+(-1)^{p_Z(p_X+p_Y)}[Z,[X,Y]]+(-1)^{p_X(p_Y+p_Z)}[Y,[Z,X]].
\label{ji}
\end{equation}

The super space of dimension (p,q) is the affine space $\R^p$ together
with a
 super algebra $\mathcal{S}^{p,q}=C^\infty(\R^p)\otimes\Lambda(\R^q)$ (the
algebra
of ``functions'' on superspace), where
$\Lambda(\R^q)=\sum_{i=0}^q\Lambda^i(\R^q)$ is  the exterior algebra of
$q$
 symbols $\theta^1,\theta^2,\dots \theta^q$. We assign grade one to the
 symbols $\theta^i$. It is clear what are the even and odd subspaces.

 An element of this superalgebra can be written as
$$
a(x,\theta)=a_0(x)+a_i(x)\theta^i+a_{i_1i_2}\theta^{i_1}\wedge
\theta^{i_2} +\cdots + a_{i_1i_2\dots i_q}\theta^{i_1}\wedge
\theta^{i_2}\cdots \wedge\theta^{i_q},
$$
where $a_{i_1i_2\dots i_j}$ is antisymmetric in all its indices.
It is a commutative superalgebra, that is,
$$
a\cdot b=(-1)^{p_ap_b}b\cdot a
$$
for homogeneous elements of degrees $p_a$ and $p_b$.

A left derivation of degree $m=0,1$ of  a super algebra is  a linear map 
$\partial^L:V\mapsto V$ 
such that 
$$
\partial^L(a\cdot b)=\partial^L(a)\cdot b +(-1)^{mp_a}a\cdot\partial^L(b).
$$
Graded left derivations form a $\Z_2$ graded vector space.
 Any linear map $L$  can be decomposed as the sum $L=L_0+L_1$,
where $L_0$ and $L_1$ are maps of degree 0 (they preserve the degree) and
1
 (they change the degree) respectively. If the superalgebra is
commutative,
 an even derivation has degree 0 as a linear map and an odd derivation 
has degree 1 as a linear map.

In the same way right derivations are defined,
$$
\partial^R(a\cdot b)=(-1)^{mp_b}\partial^R(a)\cdot b +a\cdot\partial^R(b).
$$
Notice that  derivations of degree zero are both, right and left.

A super Poisson structure on a commutative (this condition could be
relaxed,
 in particular, to introduce matrix valued superfields) super algebra is a
 super Lie algebra structure
 $\{\;,\;\}$ on it  which is also a bi-derivation with respect to the
 commutative super algebra product. More specifically, it satisfies the
following derivation property on homogeneous elements,
$$
\{a,b\cdot c\}=\{a,b\}\cdot c+(-1)^{p_ap_b}a\cdot \{b,c\},
$$
which together with the antisymmetry property (\ref{sp}) implies
$$
\{b\cdot c,a\}=b\cdot \{c,a\}+(-1)^{p_ap_c}\{b,a\}\cdot c.
$$
So, for example, if $a$ is even, $\{a,\,\cdot\,\}$ is a derivation of
degree zero, and if it is odd it is a left derivation of degree 1.

\paragraph{Example.} Consider the superalgebra   $\mathcal{S}^{p,2}$, with 
elements
$$
\Phi(x,\theta)=\Phi_0(x)+\Phi_{\alpha}(x)\theta_\alpha+
\Phi_{\alpha\beta}(x)\theta_\alpha\wedge\theta_\beta.
$$
 The derivations $\partial_i$ defined as
$$
\partial_i\Phi(x,\theta)=\partial_i\Phi_0(x)+\partial_i\Phi_{\alpha}(x)
\theta_\alpha+
\partial_i\Phi_{\alpha\beta}(x)\theta_\alpha\wedge\theta_\beta
$$
 are  even derivations. The left derivations $\partial_\alpha^L$ defined
as
$$
 \partial^L_\alpha\Phi(x,\theta)=\Phi_{\alpha}(x)+
2\Phi_{\alpha\beta}(x)\theta_\beta
$$
are odd. We can also define right derivations,
$$
 \partial^R_\alpha\Phi(x,\theta)=\Phi_{\alpha}(x)+
2\Phi_{\beta\alpha}(x)\theta_\beta
$$
Notice that $\partial^R_\alpha=\partial^L_\alpha$ on odd elements and
$\partial^R_\alpha=-\partial^L_\alpha$ on even elements. This implies
that $[\partial_\alpha^R,\partial_\beta^L]_-=0$

 These definitions can easily be extended to algebras with bigger
odd dimension in an obvious manner.

\bigskip

As an example, consider the following super Poisson bracket
\begin{equation}
\{\Phi,\Psi\}=P^{ab}\partial_a\Phi\partial_b\Psi +
P^{\alpha \beta}\partial^R_\alpha\Phi\partial^L_\beta\Psi=
P^{AB}\partial^R_A\Phi\partial^L_B\Psi. 
\label{spb}
\end{equation}
where  $P$ is a  constant matrix and satisfies the symmetry properties
$$
P^{ab}=-P^{ba},\quad P^{\alpha\beta}=P^{\beta\alpha}.
$$
It is easy to see that it satisfies the Jacobi identity (\ref{ji}).

\subsection{Super star product.}
A generalisation of the Moyal-Weyl \cite{m} deformation of
$C^\infty(\R^n)$ to $\mathcal{S}^{p,q}$ exists. This algebra structure
corresponds to the quantization of systems with both, bosonic and
fermionic degrees of freedom, and it was studied by Berezin
 as early as  in \cite{be}. There, the quantization was studied in
 terms of products of Weyl  symbols of
operators, very much in the same spirit than \cite{m}. In a language closer to
ours, it appeared  in  \cite{bf} and   in \cite{fza, fl}.

 We remind that a deformation of the commutative algebra $C^\infty(\R^p)$
is an associative
product on the space of formal series on a parameter $h$ with coefficients
in
$C^\infty(\R^n)$, that is
$C^\infty(\R^n)[[h]]=\R[[h]]\otimes C^\infty(\R^n)$. The term of first
order in $h$, antisymmetrized, is always a Poisson bracket.

We denote by $P(f\otimes g)=\{f,g\}$ a  super Poisson bracket like
(\ref{spb}),
in a space of arbitrary odd dimension. A deformation of the commutative
superalgebra $\mathcal{S}^{p,q}$ is then given by
\begin{eqnarray}
\star:\mathcal{S}^{p,q}[[h]]\otimes \mathcal{S}^{p,q}[[h]]&\longrightarrow
&\mathcal{S}^{p,q}[[h]]\nonumber\\
f\otimes g&\mapsto & e^{hP}(f\otimes g)\label{ssp}
\end{eqnarray}
where we have denoted 
$$
e^{hP}=\sum_{n=0}^\infty \frac{h^n}{n!}P^n
$$
with
$$
P^n(f\otimes g)=P^{A_1B_1}P^{A_2B_2}\cdots
P^{A_nB_n}(\partial^R_{A_1}\partial^R_{A_2}\dots
\partial^R_{A_n})f\cdot(\partial^L_{B_1}\partial^L_{B_2}
\dots\partial^L_{B_n}g).
$$
(We remind here that $P^{AB}$ is a constant matrix).
For $q=0$ this is the standard Moyal-Weyl deformation. Notice
that the first order term is exactly the super Poisson bracket.
The proof of the associativity of this product is parallel to the one
developed in \cite{bffls} for the bosonic case.

\subsection{Non formal deformation.}

We consider the associative algebra over $\R[[h]]$, $\mathcal{A}^{p,q}$, 
 generated by the symbols $X^1,\dots , X^p$,
 $\Theta^1,\dots\Theta^q$ and the relations given by the super Poisson
bracket (\ref{spb}). 

\begin{eqnarray}
&&[X^a,X^b]_{-}=hP^{ab},     
\label{ha}\\
 &&[\Theta^\alpha,\Theta^\beta]_+=hP^{\alpha\beta}.
\label{sha}
\end{eqnarray}
where $h$ is a formal parameter.  Since $X$'s and $\Theta$'s commute, it
is clear that $\mathcal{A}^{p,q}\approx U^p_h\otimes\Lambda^q_h$, where
$U_h^p$ is the associative algebra over $\R[[h]]$ generated by the symbols
$X$'s and
relations (\ref{ha}) and  $\Lambda^q_h$ is the associative algebra over
$\R[[h]]$ 
generated by the symbols $\Theta$'s and relations (\ref{sha}).
 $\mathcal{A}^{p,q}$  is isomorphic to $(\mbox{Pol}(\R^p)\otimes
\Lambda(\R^q)[[h]],\;\star\;)$, (polynomials are closed under the $\star$
operation). To prove this, we take a basis in $\mbox{Pol}(\R^p)[[h]]$,
\begin{equation}
x^{i_1}\cdot x^{i_2}\cdots x^{i_n},\qquad i_1\leq i_2\leq\cdots \leq i_n.
\label{b}
\end{equation} 
There is a $\R[[h]]$-module isomorphism
$\mbox{Sym}:\mbox{Pol}(\R^p)[[h]]\mapsto U_h^p$ mapping the elements of
the basis (\ref{b}) in the following way
\begin{eqnarray*}
&&\mbox{Sym}(x^{i_1} x^{i_2}\cdots
x^{i_n})=\frac{1}{n}\sum\limits_{\sigma\in S_n}
X^{\sigma(i_1)}\cdot X^{\sigma(i_2)}\cdots X^{\sigma(i_n)}=\\
&&\mbox{exp}({X^i\partial_i})(x^{i_1} x^{i_2}\cdots x^{i_n})|_{x^{i_k}=0},
\end{eqnarray*}
which is the usual Weyl or symmetric ordering. The  product in
$\mbox{Pol}(\R^p)[[h]]$ defined by
\begin{equation}
\mbox{Sym}^{-1}(\mbox{Sym}(f)\cdot\mbox{Sym}(g))
\label{iso}
\end{equation}
is equal to  $\star$ in (\ref{ssp}) restricted to polynomials. The
proof of this fact  is given in \cite{m} (where indeed, the argument is
extended to
$C^\infty$ functions).

Consider the basis in $\Lambda(\R^q)[[h]]$
\begin{equation}
\theta_{i_1}\wedge\theta_{i_2}\wedge\cdots \wedge \theta_{i_n},
\qquad i_1\leq i_2\leq\cdots \leq i_n.
\label{sb}
\end{equation}
(dim($\Lambda(\R^q)$)=$2^q$).
We define the isomorphism $\mbox{Sym}:\Lambda(\R^q)[[h]]\mapsto
\Lambda^q_h$ as
$$
\mbox{Sym}(\theta_{i_1}\wedge\theta_{i_2}\wedge\cdots \wedge
\theta_{i_n})=
\Theta_{i_1}\Theta_{i_2}\cdots  \Theta_{i_n}
$$
for the elements of the basis (\ref{sb}). This is the equivalent of the
Weyl ordering for odd generators. One can  see directly by inspection that
the product defined on $\Lambda(\R^q)[[h]]$ by this isomorphism is the
same than $\star$ in (\ref{ssp}) restricted to the exterior algebra. So
we can conclude that the algebra generated by $X,\Theta$ and relations
(\ref{ha}) and (\ref{sha}) is isomorphic to the $\star$-product algebra.
Given
any $\R[[h]]$-module isomorphism among  $\mbox{Pol}(\R^p)[[h]]$ and
$U_h^p$,
one can construct a star product as in (\ref{iso}). The resulting
(isomorphic) star products are called equivalent.

For polynomials, the formal parameter $h$ can be specialized to any real
value and one obtains a convergent star product. We want to look closer
to this algebra over $\R$.

By a linear change of
coordinates $P\mapsto A^TPA$, we can always bring the matrices $P^{ab}$
and
$P^{\alpha\beta}$ into a canonical form  \cite{am}, that is
\begin{equation}
P^{ab}=\begin{pmatrix} 0 & I & 0 \\
-I & 0 & 0\\
0 & 0 & 0\end{pmatrix},\qquad
P^{\alpha\beta}=\begin{pmatrix}\eta_1 & 0 & \dots & 0\\
0 &\eta_2 & \dots & 0\\
\hdotsfor{4}\\
0 & 0 &\dots & \eta_q \end{pmatrix}
\end{equation}
where $\eta_\alpha=\pm 1$ for $\alpha=1,\dots q'$ and $\eta_\alpha=0$ for
$\alpha=q'+1,\dots q$.

Denote  $q'=m+n$, where $\eta_\alpha=-1$ for $\alpha=1,\dots m$ and
$\eta_\alpha=+1$ for $\alpha=m+1,\dots m+n$. It is obvious that
$\Lambda^q_h$, evaluated for a real value of $h$ is isomorphic to the
Clifford algebra $\mathcal{C}(m,n)$ tensor product with the exterior
algebra on the remaining $q-q'$ generators, which doesn't get deformed.
(The isomorphism is given by $\gamma_\alpha=\sqrt{2h}\Theta_\alpha$).
This relation with Clifford algebras was  noticed in \cite{be}.

If $m=n$, we can make again a linear change of variables that brings
$P^{\alpha\beta}$ to the form
$$
P^{\alpha\beta}=\begin{pmatrix}0 & 1 & 0 & 0 & \dots & 0\\
                               1 & 0 & 0 & 0 & \dots & 0\\
                               0 & 0 & 0 & 1 & \dots & 0\\
                               0 & 0 & 1 & 0 & \dots & 0\\
                               \hdotsfor{6}\\
                               0 & 0 & 0 & 0 & \dots & 0 \end{pmatrix}.
$$
We have then $n$ pairs of canonically conjugate fermionic variables.
 The Poisson bracket for 
fermionic variables was first written in \cite{mi}.

\section{Formal deformations of rigid supersymmetry.}

We are interested in describing physical theories  defined on  a
deformation 
of superspace. Superfields are used as basic objects of such theories.
Mathematically,
 they are a generalisation of the superalgebra $\mathcal{S}^{p,q}$.
Consider the
trivial bundle over $\R^p$ with fibre the Grassmann or exterior algebra
$\Lambda(\epsilon_1,\dots,\epsilon_n)$. Consider the algebra of sections
on that bundle, $\Gamma^q(\R^p)$ and the tensor product
$\Phi^{p,q}_n=\Gamma^n(\R^p)\otimes\Lambda^n $.

$\Phi^{p,q}_n$ is a commutative superalgebra with the product defined as
usual
$$
(a\otimes \Psi_1)(b\otimes \Psi_2)=(-1)^{p_1p_b}ab\otimes
\Psi_1\Psi_2\qquad a,b\in \Gamma^n(\R^p),
\quad f,g\in \Lambda^n,
$$
and the left and right $\Gamma^n(\R^p)$-module structures are given by
$$
b(a\otimes \Psi)=(ba\otimes \Psi),\qquad (a\otimes
\Psi)b=(-1)^{p_bp_\Psi}(ab\otimes \Psi).
$$
The rank  ($n$) of the trivial bundle can be  chosen  
$n=\mbox{dim}\Lambda^q=2^q$. Then scalar superfields are an  even
subalgebra of $\Phi^{p,q}_n$,
 generated by elements of the form. 
$$
\Phi(x,\theta)=\Phi_0(x) +\theta_i\otimes\Phi_i(x)
+\theta_j\wedge\theta_j\otimes\Phi_{ij}+\cdots
$$
where $\Phi_{i_1i_2\dots i_k}$ are independent global sections in
$\Gamma^q(\R^p)$, antisymmetric
in the indices ${i_1i_2\dots i_k}$.

One can  extend (\ref{spb}) and (\ref{ssp}) to $\Phi^{p,q}_n\otimes
\R[[h]]$ by linearity. 
 It follows that  $(\Phi^{p,q}_n[[h]],\star)$ is a non commutative
superalgebra.

\bigskip
 
In what follows, we will restrict ourselves to four dimensional
space-time, although 
all considerations could be easily extended to other dimensions.                                                 
We consider   the  superspace associated  with the four dimensional
 N-extended Poincar\'e 
supersymmetry, with coordinates (or generators) $\{x^\mu,\theta^{\alpha
i},
\bar\theta^{\dot\alpha }_i\}$, where $\{x^\mu\}$ are the  coordinates of
ordinary four 
dimensional Minkowski space $\mathcal{M}$,   and $\{\theta^{\alpha i},
\bar\theta^{\dot\alpha }_i\}$ are Weyl spinors under the Lorentz group.
We are interested in 
deformations of this superspace such that they have an action of the
supertranslation group.
 The odd supertranslations with parameters $\epsilon^{\alpha i},\bar 
\epsilon^{\dot\alpha }_i$, act on the generators of superspace as
\begin{eqnarray}
x^\mu &\mapsto& {x'}^\mu=x^\mu +i(\theta^{\alpha
i}(\sigma^\mu)_{\alpha\dot\alpha}
\bar\epsilon^{\dot\alpha }_i-\epsilon^{\alpha
i}(\sigma^\mu)_{\alpha\dot\alpha}
\bar\theta^{\dot\alpha }_i)\nonumber\\
\theta^{\alpha i}&\mapsto& {\theta'}^{\alpha i}=\theta^{\alpha
i}+\epsilon^{\alpha i}
\nonumber\\
\bar\theta^{\dot\alpha }_i&\mapsto&{\bar{\theta'}}^{\dot\alpha
}_i=\bar\theta^{\dot\alpha }_i+\bar
\epsilon^{\dot\alpha }_i.
\label{susy}
\end{eqnarray}

By convention, we write a  scalar superfield as
$$
\Phi(x^\mu,\theta^{\alpha i},\bar\theta^{\dot\alpha }_i)=\Phi_0(x)+
 \theta^{\alpha i}\Psi_{\alpha}^i + \bar\theta^{\dot\alpha }_i
\bar\Sigma_{\dot \alpha i}+\theta^{\alpha i}\theta^{\beta j}
\Psi_{\alpha\beta}^{ij} +\cdots,
$$
where we have dropped the symbols ``$\wedge$'' and ``$\otimes$''.
Let $g(\epsilon)$ be a super translation  as in (\ref{susy}). The action 
of $g$ on superfields is given by 
$(g^{-1}\Phi)(x,\theta,\bar\theta)=\Phi(x',\theta',\bar\theta')$. 
We require that the super translation  group acts as a group of
 automorphisms on the deformed algebra, that is
$$
g(\Phi_1\star \Phi_2)=(g\Phi_1)\star (g\Phi_2).
$$
It is convenient to introduce the right and left odd derivations called
super covariant derivatives, 
\begin{eqnarray*}
{D^{R,L}}_{\alpha i}&=&{\partial^{R,L}}_{\alpha i} +
(i\sigma^\mu_{\alpha\dot\alpha}\bar \theta^{\dot\alpha
}_i)\partial_\mu)^{R,L},
\\
{ {\bar D}^{R,Li} }_{\dot\alpha}&=&-{ {\bar \partial}^{R,Li}
}_{\dot\alpha}-
(i\theta^{\alpha i}\sigma^\mu_{\alpha\dot\alpha}\partial_\mu)^{R,L}.
\end{eqnarray*}
They have the property that
$$
{D^{R,L}}_{\alpha i}(g\Phi)=g({D^{R,L}}_{\alpha i}\Phi).
$$
and the same for ${ {\bar D}^{R,Li} }_{\dot\alpha}$. We define a Poisson
bracket 
\begin{equation}
\{\Phi,\Psi\}=P^{\mu\nu}\partial_\mu\Phi\partial_\nu\Psi +P^{\alpha i\beta
j}
{D^{R}}_{\alpha i}\Phi{D^{L}}_{\beta j}\Psi.
\label{spb2}
\end{equation}
The crucial properties are that $[{D^{R}}_{\alpha i},{D^{L}}_{\beta
j}]_-=0$ and that
$D^{R,L}_{\alpha i}P^{AB}=0$ ($A=\mu,\{\alpha\,i\}$),  so one 
can again construct a Weyl-Moyal star product as in  (\ref{ssp}), which
will 
also be covariant under the supertranslation group.  One could also extend
(\ref{spb2}) 
by using  $N-k$ $D$'s and $k$ $\bar D$'s $k=1,\dots N$, which can be taken
to  anticommute. 
 Notice that this Poisson structure
is   degenerate in the  space of odd variables.

A chiral field \cite{fwz} satisfies the constraint ${ {\bar
D}^{R,Li}}_{\dot\alpha}\Phi=0$. 
The solution of this equation can be written (after a change of variables)
as 
$\Phi(x,\theta)$. Chiral superfields are a subalgebra under  ordinary
multiplication,
but they are not  closed under the star product
defined with (\ref{spb2}) unless $P^{\alpha i\beta j}=0$. 

\bigskip

When considering extended supersymmetry this notion of chirality can be 
generalised.  The R-symmetry group
  U($N$) acts   by automorphisms on the super 
Poincar\'e algebra, leaving invariant the even generators. For $N>1$,
one can take the direct product of the Minkowski space with the flag
manifold
SU($N$)/U(1)$^{N-1}$, and consider a  supermanifold structure of odd 
dimension 
$4N$ on it. This is constructed by taking the quotient 
$(\mathcal{L}\otimes\mbox{SU}(N))\otimes_s\mathcal{ST}/\mathcal{L}\otimes
\mbox{U(1)}^{N-1}$,
where $\mathcal{L}$ is the Lorentz group and $\mathcal{ST}$ is the
supertranslation group.
 It it is called harmonic superspace \cite{gikos}.
 The algebra of global sections (or functions) on the
resulting supermanifold is isomorphic to 
 \begin{equation}
C^\infty(\mathcal{M}\times
 \hbox{SU}(N)/U(1)^{N-1})\otimes \Lambda^{4N}.
\label{hs}
\end{equation}
This isomorphism is not canonical,
 since it is not preserved by supersymmetry transformations, but it is
preserved 
by the action of the R-symmetry group.

Let us  denote the coordinates
 in an open set as $\{x^\mu, u,\theta^{\alpha i}, \bar\theta^{\dot\alpha
}_i\}$
 where $u$ is a unitary matrix or coset representative of
SU($N$)/U(1)$^{N-1}$. With the coset 
representatives one can define  rotated covariant derivatives
$$
\mathcal{D}_{\alpha I }=u_I^iD_{\alpha i}, \qquad 
\bar{\mathcal{D}}_{\dot\alpha}^I=u_I^i\bar{D}_{\dot\alpha}^i.
$$
 The advantage of such 
formulation is that the notion of chiral superfield can be generalised 
by imposing the following R-symmetry covariant  constraints on the
superfields $\Phi(x,u,\theta,
\bar\theta)$,
$$
\mathcal{D}_{\alpha 1}\Phi=\cdots=\mathcal{D}_{\alpha k}\Phi=0=
\bar{\mathcal{D}}_{\dot \alpha}^{k+1}\Phi=\dots=\bar{\mathcal{D}}_{\alpha
}^N\Phi.
$$
(no superscript will mean that we are considering a left derivative).
The solution of these constraints can be expressed as superfields that do
 not depend on  $k$ $\theta$'s and
 $N-k$ $\bar\theta$'s  ($k=0,N$ being the usual chiral and antichiral 
superfields). These fields have been called   
``Grassmann analytic'' in the literature \cite{fs} and, as chiral 
superfields, they form a subalgebra.

One can consider deformations of this supermanifold for a given super 
Poisson structure. In particular one can 
consider a deformation affecting only  the first factor in (\ref{hs}). 
Any deformation of this form will have the supersymmetry algebra as an 
algebra of derivations. As 
the simplest case, let us take a non trivial Poisson bracket only in the
 directions of $\mathcal{M}$,
$$iP=iP^{\mu \nu}\frac{\partial}{ \partial x^\mu}\otimes\frac{\partial}{
\partial x^\nu}
$$
that is, the Poisson bracket of two (complex) superfields is
$$
\{\Phi_1,\Phi_2\}=iP^{\mu \nu}\frac{\partial\Phi_1}{\partial x^\mu}
\frac{\partial\Phi_2}{\partial x^\nu}
$$
where $P^{\mu\nu}$ is an arbitrary constant antisymmetric matrix and 
$\Phi_i(x,u,\theta)$ arbitrary superfields. Then, the 
Weyl-Moyal  star product on the algebra of superfields is given  by
\begin{equation}
\Phi_1\star\Phi_2= \exp(iP)(\Phi_1\otimes\Phi_2).
\label{gas}\end{equation}

It is clear from this expression that Grassmann analytic superfields are
 closed under the star product (\ref{gas}), as chiral superfields are.

\section{Non commutative Wess-Zumino model.}

The simplest example of an   N=1  supersymmetric 
field theory is the Wess-Zumino model, whose action is
$$
\int d^4xd^\theta d^2\bar\theta\;  \Phi\bar\Phi+\int d^4x\;(\int
d^2\theta\; (\frac{m}{ 2}\Phi^2 +
\frac{g}{ 3}\Phi^3) +\mbox{c. c.}),
$$ 
where he chiral superfield $\Phi$ has the expansion
$$
\Phi=A(y)+\sqrt{2}\theta\psi(y) +\theta\theta F(y),
$$
where $y=x+i\theta\sigma\bar\theta$.
A formal deformation of this action can be written using the star product 
defined above (\ref{gas}),
\begin{equation}
\int d^4xd^2\theta d^2\bar\theta\; \Phi\bar\Phi+\int d^4x\;(\int d^2\theta
\;(\frac{m}{2}\Phi^2 +
\frac{g}{3}\Phi^{\star 3}) +\mbox{c. c.}).
\label{incw}
\end{equation} 
where $\Phi^{\star n}=\Phi\star\Phi\cdots(n)\cdots\star\Phi$. This model
was also considered in \cite{cr, gp}.
This Lagrangian is unique  (for every star product)
 as a consequence of the fact that
\begin{equation}
\int d^4x\;A\star B=\int d^4x\;A B=\int d^4x\;B\star A
\label{ip1}
\end{equation} 
and
\begin{equation}
\int d^4x\;A_1\star\cdots \star  A_n=\int d^4x\;A_{\sigma(1)}\star\cdots
\star A_{\sigma(n)}
\label{ip2}
\end{equation}
where $\sigma$ is a cyclic permutation of $(1,\dots, n)$.
Notice also that the above action is real, since
$$
\overline{A\star B} =\bar B\star \bar A.
$$

As a consequence  of (\ref{ip1}) and (\ref{ip2}), the auxiliary field $F$
satisfies pure algebraic
 equations 
$$
F=-m\bar A-g\bar A\star\bar A
$$
so the component form of the Lagrangian is
\begin{eqnarray}
&& i\partial_\mu\bar\psi\bar\sigma^\mu\psi +\bar A\partial_\mu\partial^\mu
  A-\frac{1}{ 2}m(\psi\psi +\bar \psi \bar\psi) -m^2\bar A A
-g(A(\psi\star \psi) +\nonumber\\&&\bar A(\bar \psi\star \bar\psi))
-mg(A(\bar A\star \bar A )+ \bar A(A\star A))
-g^2(A\star A)(\bar A\star \bar A).
\label{cl}
\end{eqnarray}

The non deformed Wess-Zumino model is a renormalizable field theory which
only requires 
a (logarithmically  divergent) wave function renormalization \cite{wz,
iz}. 
This is due to supersymmetric non renormalization theorems of chiral terms
\cite{fiz}.

The deformed Wess-Zumino model is the supersymmetric extension of the
$\phi^4$ theory 
considered in \cite{mrs} where the model was proven to 
be ``renormalizable''  in some extended sense.
Consequently, we expect that its supersymmetric extension is also
``renormalizable''. Moreover,
since the interactions are purely chiral, no quadratic divergences  appear
and then the UV/IR
connection induced by extra poles in the propagator \cite{mrs} does not
appear in this model \cite{cr}. Additional aspects of the UV/IR
connection in non commutative supersymmetric models are discussed in
\cite{gp, ss}.

The non deformed Lagrangian has a quartic interaction that is invariant
under a local
 U(1) symmetry. This invariance is inherited from the superconformal
symmetry 
present in the model when 
$m=0$. The deformed Lagrangian only preserves the global U(1) invariance. 
It is interesting to observe that there is another possible quartic term
 \begin{equation}
(\bar A\star  A)^2,
\label{asq}
\end{equation}
which is  invariant
under a non commutative local U(1)-symmetry \cite{ybk}. Supersymmetry
picks 
the first choice without any contradiction because the R-symmetry is not
deformed.

Incidentally, it was also  shown that the pure bosonic theory of a complex
 scalar field $A$ with the quartic
interaction as given in Lagrangian (\ref{cl}) is not renormalizable unlike
the theory
with the quadratic invariant (\ref{asq}). This is not a contradiction
because in the Wess-Zumino model the additional interactions due to
supersymmetry are
 responsible for the cancellation of dangerous divergences (in particular
quadratic divergences).

As a side remark, we note that while the interaction $\bar A\star\bar
A\star A\star A$ is
typical of an $F$-term, the other possibility $\bar A\star\ A\star\bar
A\star A$ is typical
 of a $D$-term, so we expect the latter to occur in the deformed version
of supersymmetric Q.E.D.
Even more, both quartic terms occur and in fact are related one to another
when $N$-extended
supersymmetry is present. This will be the case in the deformed version of
$N=2,4$ super
 Yang-Mills theories, which in addition require a deformation of the gauge
symmetry.

\section{Non commutative rank 1 gauge theory in superspace.}

In this section we introduce a deformation of an abelian gauge theory in
superspace
\cite{wz2, fz}.
The gauge group is a group of formal series in a parameter with
coefficients which are 
 chiral superfields ($U$, with $\bar D_{\dot\alpha}U=0$).
The multiplication law is given by the star product in  (\ref{gas})
$$
U_1\star U_2=U_3,
$$
 which preserves chirality.
We will denote by $U^{\star -1}$ the inverse with respect to the star
product,
$$
U\star U^{\star -1}=U^{\star -1}\star U=1.
$$
Notice that for $\theta=\bar\theta=0$, the gauge parameter is a complex
function, so 
the gauge group is  the complexification of U(1).  We can write an element
$U$ as
$$ 
U=e^{\star i\Lambda}=\sum_{n=0}^\infty \frac{1}{n!}(i\Lambda)^{\star n},
$$
and then
$$
U^{\star -1}=e^{\star - i\Lambda},\qquad U^\dagger=e^{\star - i\bar
\Lambda},\qquad 
{U^\dagger}^{\star -1}=e^{\star  i\bar \Lambda}.
$$
We introduce a connection superfield $V$ \cite{fz}  which  transforms
under the gauge group as
\begin{eqnarray*}
&&e^{\star V}\mapsto U^\dagger\star e^{\star V}\star U\\
&&e^{\star -V}\mapsto U^{\star -1}\star e^{\star - V}\star
{U^\dagger}^{\star -1}.
\end{eqnarray*}
The non commutative field strength
$$
W_\alpha={\bar D}^2(e^{\star -V}\star D_\alpha e^{\star V}),\qquad \bar
D_{\dot \alpha}W_\alpha=0,
$$
transforms as
$$
W_\alpha\mapsto U^{\star -1}\star W_\alpha \star U.
$$
The action
\begin{equation}
\mathcal{S}_{NCYM}=\int d^4x\;(\int d^2\theta\; W_\alpha\star W^\alpha +
\mbox{c.c.})
\label{ncym}
\end{equation}
defines the non commutative rank 1 gauge theory. It is gauge invariant as
a consequence of (\ref{ip1}).
 If we set the gaugino $\lambda$ and the auxiliary field $D$ to zero,
this action reduces to the bosonic non commutative action considered in
\cite{sw}. Note also
that the equation of motion of the auxiliary field is $D=0$.

The infinitesimal gauge transformation of the connection superfield is an
infinite power series 
with terms  of the type $V^{\star n}$. To first order in $V$ it is
$$
\delta V=i(\Lambda-\bar \Lambda)-\frac{1}{2}i[(\Lambda+\bar \Lambda)\star
V-V\star(\Lambda +\bar\Lambda)].
$$
This is actually the transformation in the Wess-Zumino gauge ($V^{\star
3}=0$) 
\cite{wz2, fz}. In this gauge the field
strength becomes 
$$
W_\alpha=D_\alpha V -\frac{1}{2}(V\star D_\alpha V-D_\alpha V\star V).
$$
Since the Wess-Zumino gauge depends on the deformation parameter $P$, the
modified 
supersymmetry transformations which preserve this gauge will also depend
on $P$. Indeed,
the gaugino transformation contains the   two-form field strength $F=dA
+iA\star A$. Also, the 
supersymmetry transformation of the auxiliary field $D$ contains the
covariant 
derivative of the gaugino $\nabla \lambda =d\lambda + i(A\star \lambda
-\lambda\star A)$.

It is worth noticing that the action in (\ref{ncym})  is invariant under a
non linearly realised 
supersymmetry transformation 
$$
\delta W_\alpha =\eta_\alpha
$$
where $\eta_\alpha$ is a constant, anticommuting spinor. This leads to the
interpretation of
 a non commutative Yang-Mills theory as a Goldstone action of partial
breaking of supersymmetry.

We may now couple this action to matter chiral multiplets $S_i$. This can
be done in two different ways,
depending whether we introduce adjoint matter $S\mapsto U^{\star -1}\star
S\star U$ (which is
neutral in the commutative limit), or charged matter $S\mapsto S\star U$.
In the first case the non commutative gauge invariant coupling is
\begin{equation}
\int d^4x\int d^2\theta d^2\bar \theta\; S\star e^{\star -V}\star\bar
S\star e^{\star V}.
\label{ci}
\end{equation}
Note that we can add to the action any chiral interaction such as
(\ref{incw}), which will be 
automatically gauge invariant.

 If we now consider   a single chiral multiplet and a vector   multiplet,
then the sum of the two actions
(\ref{ncym}) and (\ref{ci}) is known to have in the commutative limit
$N=2$ supersymmetry. Therefore, 
following the discussion in section 3, the deformed theory is the first
example of 
deformed theory with $N=2$ supersymmetry. This theory could in fact be
reformulated using harmonic superspace
which is the natural set up for $N=2$ Yang-mills theories \cite{gikos}.

If we introduce three chiral adjoint multiplets $S_i,\; i=1,\dots 3$
with an additional  self coupling
$$
\int d^2\theta \;\epsilon^{ijk}S_i\star S_j\star S_k +\mbox{c. c. },
$$
(which vanishes in the commutative limit) we obtain a deformation of
$N=4$, rank 
1 supersymmetric Yang-Mills theory, which could also be reformulated in
harmonic superspace \cite{ka, hw}.
The Yang-Mills field is then a Grassmann analytic function \cite{gikos,
fs} 
and it is important that the star product 
(\ref{gas}) preserves Grassmann analyticity.

 $N$-extended non commutative rank 1 gauge theories are invariant under
$N$ non linearly 
realized supersymmetries, corresponding to constant shifts of the gauginos
by anticommuting
 parameters. This is due to the fact that the  cubic interactions involving
gauginos, under such 
transformation, vary into a Moyal bracket (antisymmetrized star product),
which is a total 
space-time derivative. This brings more evidence to the fact that such
theories 
are closely connected to world volume brane theories which, as a
microscopic 
description of 1/2 BPS states, have $2N$ supersymmetries, with half of
them 
non linearly realized in the  spontaneously broken phase \cite{bg, t}.

For a charged matter field, the gauge invariant non commutative action is
$$
\int d^4x \int d^2\theta d^2\bar\theta\; S\star e^{\star -V}\star\bar S.
$$

\bigskip

It was shown in \cite{sw} that the phase space of a non commutative
Yang-Mills theory 
can be mapped to the phase space of an ordinary Yang-Mills theory by a
``change of variables''
realised in the following way: the gauge potential of the ordinary gauge
group 
$A$ is mapped into the gauge potential $\hat A (A)$ 
of the non commutative (deformed) gauge group , while the gauge group
parameter $\lambda$
 is mapped into the 
noncommutative gauge group parameter $\hat \lambda(\lambda, A)$ in such a
way that the 
respective gauge transformations satisfy
$$
\hat A(A) +\hat\delta_{\hat\lambda}\hat A(A)=\hat A(A+\delta_\lambda A).
$$
For rank one and to first order in $P$, the solution to this differential
equation is \cite{sw}
\begin{eqnarray}
&&\hat A_\mu(A)=A_\mu -\frac{1}{2}P^{\rho\sigma}A_\rho(\partial_\sigma
A_\mu +F_{\sigma\mu})\nonumber\\
&&\hat\lambda(\lambda, A)=\lambda+\frac{1}{4}P^{\mu\nu}\partial_\mu\lambda
A_\nu.
\label{cv}
\end{eqnarray} 

These transformations can be supersymmetrized as follows. The  non
commutative  gauge connection 
and gauge parameter superfields will be now denoted by $\hat V$ and $\hat
\Lambda$, while we 
will reserve the notation $V$ and $\Lambda$ for their ordinary
counterparts.
The transformations then read,
\begin{eqnarray}
&&\hat V(V)=V +aP^{\mu\nu}\partial_\mu V\nabla_\nu V + (b
P^{\alpha\beta}D_\alpha VW_\beta +
 \mbox{c. c.}) +\nonumber\\
&&(c P^{\alpha\beta}VD_\alpha W_\beta + \mbox{c. c.}),\nonumber\\
&& \hat\Lambda(\Lambda, V)= \Lambda +d{\bar D}^2(P^{\alpha\beta}D_\alpha
\Lambda D_\beta V),
\label{scv}
\end{eqnarray}
where 
\begin{eqnarray*}
P^{\alpha \beta}&=&(\sigma_{\mu\nu})^{\alpha\beta}P^{\mu\nu}, \qquad
(\mbox{symmetric in }(\alpha,\beta)),\\
  W_\alpha&=&{\bar D}^2D_\alpha V,\\
\nabla_\nu&=&(\sigma_\nu)^{\alpha\dot\alpha}[D_\alpha,D_{\dot\alpha}].
\end{eqnarray*}
$a,b,c,d$ are numerical coefficients, which are uniquely fixed in order to
reproduce (\ref{cv}).
We also want to note that $\hat \Lambda=\Lambda $ for
$D_\alpha\Lambda=0$, which is required
by consistency. Analogously, $\hat V=V$ for constant $V$.

 The above results can be easily generalised to non commutative super
Yang-Mills theories of 
arbitrary rank.

\section{The $\alpha'\mapsto 0$ limit of supersymmetric Born-Infeld action
and deformed
 U(1) gauge theory.}

In this section we will present the supersymmetric version of the
$\alpha'\mapsto 0 $
 limit of the Born-Infeld action when a $B$-field is turned on. This
action is supposed 
to describe the deformed version of supersymmetric U(1) gauge theory (in
the limit
 $\alpha'=\mathcal{O}(\epsilon^\frac{1}{2})\mapsto 0$ and slowly varying
fields) \cite{sw},  where the 
constant field $B^{\mu\nu}$ is related to the Poisson bivector by 
$B^{\mu\nu}={P^{-1}}^{\mu\nu}$ as in (\ref{gas}).

Let us first remind the expression obtained in the bosonic case. The
Lagrangian is given by
\begin{equation}
\mathcal{L}_{BI}=\sqrt{\det(\epsilon^\frac{1}{2}+F)} =\sqrt{\epsilon^2
+\frac{\epsilon}{2}F^2 +
 \frac{1}{16}(F\tilde F)^2}.\label{obi}
\end{equation}
To order $\epsilon$, the Lagrangian is readily seen to be
$$
\frac{1}{4}|F\tilde F| +\epsilon\frac{F^2 }{|F\tilde F|}.
$$
To obtain the  supersymmetric Born-Infeld action, we will use the
following identity
\cite{cf},
$$
\sqrt{X^2-Y}=X+ Y\frac{\sqrt{X^2-Y}-X}{Y}=X-\frac{Y}{\sqrt{X^2-Y}+X},
$$
where 
$$
X=\epsilon +\frac{1}{4}F^2,\qquad Y=\frac{1}{16}\big((F^2)^2-(F\tilde
F)^2\bigr).
$$
Denoting by 
$$
F_\pm=\frac{1}{2}(F\pm\tilde F)
$$
the self dual and anti self dual combinations of $F$ in 
an Euclidean metric, we also have 
$$
F_\pm^2=\frac{1}{2}(F^2\pm F\tilde F),\qquad F_+^2F_-^2=\frac{1}{4}\bigl(
(F^2)^2-(F\tilde F)^2\bigr).
$$
We consider a chiral spinor superfield  $W_\alpha$ $(\bar
D_{\dot\alpha}W_\alpha=0)$, and 
the chiral scalar superfield $T$
$$
T=\bar D\bar D{\bar W}^2=-\frac{1}{2} F_-^2 +\cdots
$$
We promote $X$ and $Y$ to superfields 
$$
X=\epsilon-\frac{1}{2}(T+\bar T),\quad Y=T\bar T.
$$
The supersymmetric Born-Infeld action is 
\begin{eqnarray}
\mathcal{L}_{SBI}&=&-\frac{1}{2}\int d^2\theta\; W^2-\frac{1}{2}\int
d^2\bar\theta\;{\bar W}^2-
\int d^2\theta d^2\bar \theta \;\frac{{\bar W}^2W^2}{\sqrt{X^2-Y} +X}.
\end{eqnarray}
One has that
$$
\sqrt{X^2-Y}=\sqrt{\epsilon^2-\epsilon(T+\bar T) +\frac{1}{4}(T-\bar
T)^2}.
$$
The order 0 in $\epsilon$ in the above expression is the square root of a
square,
so it should be understood as
\begin{equation}
\sqrt{\frac{1}{4}(T-\bar T)^2}=\pm\frac{1}{2}(T-\bar T)
\label{pms}
\end{equation}
depending if
$$
\frac{1}{2}(T-\bar T)|_{\theta=0}=\frac{1}{4}F\tilde F
$$
is grater or less than zero.
So the $\epsilon=0$ term of $\mathcal{L}_{SBI}$ is
$$
\mp\frac{1}{2}(\int d^2\theta W^2-\int d^2\bar\theta{\bar
W}^2)=\frac{1}{4}|F\tilde F|+\cdots
$$
For the order $\epsilon$ term one gets (in the case with + sign in
(\ref{pms}))
$$
2\epsilon \int d^2\theta d^2\bar\theta\; \frac{W^2{\bar W}^2}{D^2W^2
(D^2W^2-{\bar D}^2{\bar W}^2)}.
$$
(for the other case in (\ref{pms}) we exchange $W$ by $\bar W$ and $D$
by $\bar D$).

Finally we get for the $\mathcal{O}(\epsilon)$ in $\mathcal{L}_{SBI}$
\begin{eqnarray}
&&\pm\epsilon\int d^2\theta d^2\bar \theta\; W^2{\bar W}^2(\frac{1}{D^2W}+
\frac{1}{{\bar  D}^2{\bar W}^2})\frac{1}{(D^2W^2-{\bar D}^2{\bar
W}^2)}-\nonumber\\&&
\epsilon\int d^2\theta d^2\bar\theta\; \frac{W^2{\bar W}^2}{(D^2W^2{\bar
D}^2{\bar W}^2)} 
= \epsilon(\frac{F^2}{|F\tilde F|}-1) +\cdots.
\label{fwg}
\end{eqnarray}
Note that the last term in (\ref{fwg}) corresponds to a shift by $\epsilon$ in
the original
Born-Infeld action (\ref{obi}),
$$
\mathcal{L}=\sqrt{\det(\epsilon^\frac{1}{2}+F)} -\epsilon.
$$

When the $B$ field is turned on ($F\mapsto F+B$ in the bosonic action),
 the superfield strength $W_\alpha=\bar D\bar DD_\alpha V$ 
($V$ is the gauge superfield) is shifted into $W_\alpha-L_\alpha$
\cite{flz}, where $L_\alpha$ is the 
spinor chiral superfield containing the $B$ field in its
$\theta$-component,
$$
L_\alpha=\theta^\beta(\sigma^{\mu\nu}_{\alpha\beta}B_{\mu\nu}
+\epsilon_{\alpha\beta}\phi) + 
\theta^2\chi_\alpha,
$$
where we used the fact that the combination $W_\alpha-L_\alpha$ is
invariant under  the 
superspace gauge transformation 
\begin{eqnarray*}
V&\mapsto & V+U\\
L_\alpha &\mapsto &L_\alpha +{\bar D}^2D_\alpha U.
\end{eqnarray*}
where $U$ is an arbitrary real scalar superfield.
If we want now to compute  the supersymmetric Born-Infeld action in the
$\epsilon \mapsto 0$
limit with a constant $B$ field, it is then sufficient to set
$\phi=\chi_\alpha=0$, 
replace $W_\alpha$ by $W_\alpha
-\theta_\beta\sigma^{\mu\nu}_{\alpha\beta}B_{\mu\nu}$ and 
then use (\ref{fwg}).

The $\mathcal{O}(\epsilon)$ supersymmetric version of the Born-Infeld
bosonic Lagrangian,
$$
\frac{F^2}{|F\tilde F|}
$$
has a straightforward generalisation to the case of extended supersymmetry
as a full superspace 
integral \footnote{We observe that such generalizations are not unique
unless we impose additional
requirements on the Born-Infeld action such as electromagnetic
duality invariance for its equations of motion \cite{kt}.}. For $N=2$ theories we have 
$$
\mathcal{L}_{SBI}(N=2)=\int d^4\theta d^4\bar\theta\;\frac{W^2{\bar
W}^2}{D^4W^2-{\bar D}^4{\bar  W}^2}(
\frac{1}{D^4W^2}+\frac{1}{{\bar D}^4{\bar W}^2})
$$
where $W$ is the $N=2$ chiral superfield strength. This is in fact the
$\alpha'\mapsto 0$ limit of the 
$N=2$ supersymmetric Born-Infeld action \cite{k, t}.

For $N=4$ we may write an on-shell superspace action \cite{hst}
$$
\mathcal{L}_{SBI}(N=4)=\int d^8\theta d^8\bar\theta\;
\frac{W^{4(0,4,0)}W^{4(0,4,0)}|_{\mbox{singlet}}}
{F_+^2-F_-^2}(\frac{1}{F_+^4F_-^2} +\frac{1}{F_-^4F_+^2}),
$$
where the $N=4$ superfield strength $W^{ij}=-W^{ji}$ satisfies the
following constraints,
\begin{eqnarray*}
&&W^{ij}=\frac{1}{2}\epsilon^{ijkl}{\bar W}^{kl}\\
&&{\bar
D}_{i\dot\alpha}W^{jk}=\frac{1}{3}(\delta_i^jW^{lk}-\delta_i^k{\bar
D}_{l\dot\alpha}W^{lj}),\\
&&D_\alpha^iW^{jk} +D_\alpha^jW^{ik}=0,
\end{eqnarray*}
and
$$
F_+^2=D^4_{0,2,0}W^{2(0,2,0)}|_{\mbox{singlet}},\qquad F_-^2={\bar
D}^4_{0,2,0}W^{2(0,2,0)}|_{\mbox{singlet}}.
$$
The indices $(a,b,c)$ refer to the SU(4) Dynkin labels, and ``singlet''
means a projection on SU(4)
 invariant combinations.

It is a challenging problem to show that the above actions should
reproduce a deformation 
of the supersymmetric U(1) gauge theory.

\vfill\eject
\noindent{\Large \bf Acknowledgments}

\bigskip

We would like to thank Yaron Oz for a useful discussion.
M. A. Ll. is thankful to the Theory Division of CERN for its kind
hospitality. 
The work of S. F. has been supported in part by the European Commission
TMR program
ERBFMRX-CT96-0045 (Laboratori Nazionali di Frascati, INFN) and by DOE
grant DE-
FG03-91ER40662.

\end{document}